%% file: main.tex
\documentclass[sigconf, authorversion]{acmart}
\input{dlabmacros.tex}
\usepackage{mathrsfs}
\usepackage{multirow}
\usepackage{enumitem}
\usepackage{pifont}
\newcommand{\imark}{\ding{74}}
\newcommand{\cmark}{\ding{51}}%
\newcommand{\xmark}{\ding{55}}%

\newcommand{\rfpp}{{RfPP}}



\usepackage{subcaption}

\usepackage{tikz}
\usetikzlibrary{positioning, calc, shapes.geometric, shapes, shapes.multipart, arrows.meta, arrows, decorations.markings, external, trees}
\tikzset{
node distance=0.5cm, 
}
\tikzstyle{Arrow} = [
	thick, 
	decoration={
		markings,
		mark=at position 1 with {
			\arrow[thick]{latex}
			}
		}, 
	shorten >= 0.5pt, preaction = {decorate},
	]

\AtBeginDocument{%
  \providecommand\BibTeX{{
    \normalfont B\kern-0.5em{\scshape i\kern-0.25em b}\kern-0.8em\TeX}}}

\copyrightyear{2023}
\acmYear{2023}
\setcopyright{rightsretained}
\acmConference[Under Review]{Under Review}{2024}{Anonymous Location}
\acmBooktitle{Under Review}
\acmPrice{15.00}
\acmDOI{xx.xxxx}
\acmISBN{xxx}

\begin{document}

\title[Protection from Evil and Good: The Differential Effects of Page Protection on Wikipedia Article Quality]{Protection from Evil and Good: The Differential Effects of \\Page Protection on Wikipedia Article Quality}


\author{Thorsten Ruprechter}
\affiliation{%
  \institution{TU Graz}
    \country{Graz, Austria}
}
\email{ruprechter@tugraz.at}
\author{Manoel Horta Ribeiro}
\affiliation{%
  \institution{EPFL}
    \country{Lausanne, Switzerland}
}
\email{manoel.hortaribeiro@epfl.ch}

\author{Robert West}
\affiliation{%
  \institution{EPFL}
  \country{Lausanne, Switzerland}
}
\email{robert.west@epfl.ch}

\author{Denis Helic}
\affiliation{%
  \institution{Modul University Vienna}
  \country{Vienna, Austria}
}
\email{denis.helic@modul.ac.at}

\renewcommand{\shortauthors}{Ruprechter et al.}

\begin{abstract}
Wikipedia, the Web's largest encyclopedia, frequently faces content disputes or malicious users seeking to subvert its integrity.
Administrators can mitigate such disruptions by enforcing ``page protection'' that selectively limits contributions to specific articles to help prevent the degradation of content.
However, this practice contradicts one of Wikipedia's fundamental principles---that it is open to all contributors---and may hinder further improvement of the encyclopedia.
In this paper, we examine the effect of page protection on article quality to better understand whether and when page protections are warranted.
Using decade-long data on page protections from the English Wikipedia, we conduct a quasi-experimental study analyzing pages that received ``requests for page protection''---written appeals submitted by Wikipedia editors to administrators to impose page protections. 
We match pages that indeed received page protection with similar pages that did not and quantify the causal effect of the interventions on a well-established measure of article quality.
Our findings indicate that the effect of page protection on article quality depends on the characteristics of the page prior to the intervention: high-quality articles are affected positively as opposed to low-quality articles that are impacted negatively. 
Subsequent analysis suggests that high-quality articles degrade when left unprotected, whereas low-quality articles improve.
Overall, with our study, we outline page protections on Wikipedia and inform best practices on whether and when to protect an article.
\end{abstract}


\maketitle

\begin{figure}[t]
	\centering
	\begin{subfigure}[b]{.72\columnwidth}
		\centering
		\includegraphics[width=\textwidth]{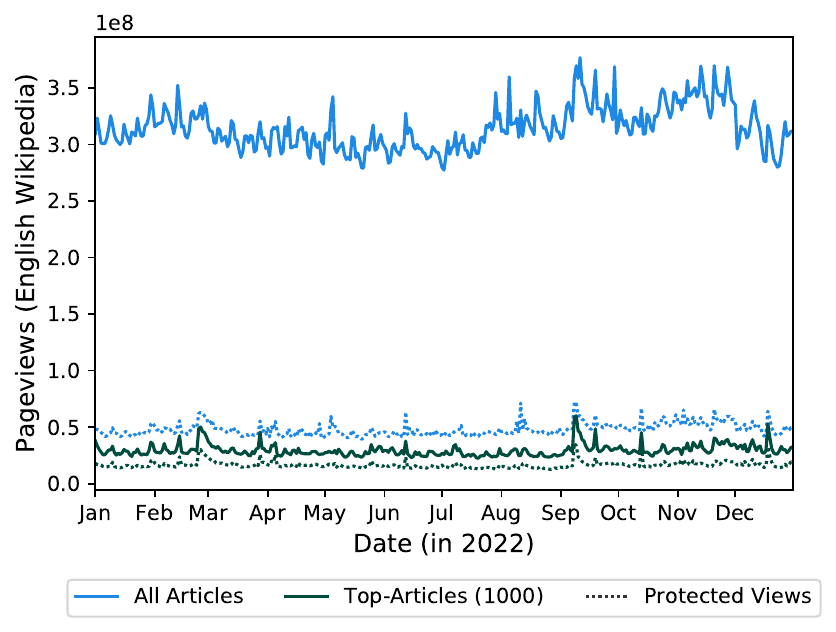}
		\caption{Time series of pageviews in 2022.}
		\label{fig:pp_views_single}
	\end{subfigure}
	\begin{subfigure}[b]{.265\columnwidth}
		\centering
		 \raisebox{-20mm} {\includegraphics[width=\textwidth]{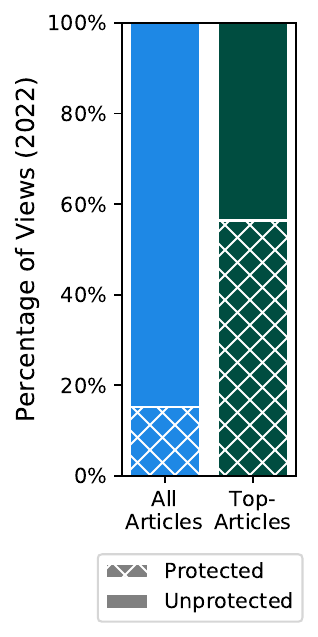}}
		\caption{Aggregated (\%).}
		\label{fig:pp_views_bar}
	\end{subfigure}
	\caption{
            Page protected articles correspond to a substantial fraction of page views Wikipedia receives.
            In (a), we show the daily number of views in 2022 that went towards: 
            all English Wikipedia articles (blue), 
            the $\mathbf{1{,}000}$ most-viewed articles per day (dark green),
            and page protected articles in both of these sets (dotted lines).
            From (b), we derive that, over the whole year, approximately $\mathbf{15\%}$ of overall views and $\mathbf{56\%}$ of views to the $\mathbf{1{,}000}$ top articles went to protected articles.
		}
		\label{fig:prot_views}
\end{figure}


\section{Introduction}

Wikipedia, one of the largest repositories of knowledge on the Web, serves the diverse information needs of users around the world~\cite{Lemmerich2019}.
From general topics, such as science and history~\cite{singer2017we}, to events, such as pandemics~\cite{ruprechter2021volunteer} and natural disasters~\cite{lorini2020uneven}, the world turns to the online encyclopedia for fast, reliable, and free information.
Wikipedia is made possible due to the work of volunteers (\textit{Wikipedians}) who create new articles and curate existing content. 
At the core of Wikipedia's philosophy is its openness: ``Anyone---including you---can become a Wikipedian by boldly making changes when they find something that can be added or improved~\cite{wkpends}.''

While most contributions to Wikipedia are constructive, good-faith editors may find themselves embroiled in edit wars about verbiage or conflicts about the validity of information~\cite{faulkner2012etiquette, sumi2011edit}; and malicious users might vandalize pages or manipulate the tone of an article through sockpuppetry~\cite{kittur2007he, kumar2017army}, undermining the development of content and the integrity of knowledge on Wikipedia~\cite{aragon2021preliminary}.

Regardless of users' intent, when disruptive editing patterns emerge, Wikipedia admins frequently intervene using ``page protection'' to safeguard them against further damage~\cite{burke2008mopping, hill2015page}.
This mechanism restricts who can edit an article temporarily (e.g., one week) or indefinitely and can be triggered by admins, either at their own will or when fulfilling user requests to protect a page~\cite{rfppGeneral}.
Although less than $1\%$ of pages in the English Wikipedia are protected at any given time~\cite{hill2015page, spezzano2019detecting}, protected articles receive a disproportionate amount of views (e.g., 15\% of all views in 2022; see Fig.~\ref{fig:prot_views}).

While page protections can be effective in curbing disruptive behavior, they are at odds with Wikipedia's openness principle, and thus, internal policies advocate using them scarcely~\cite{wikiPPPolicy}.
They can deter newcomers~\cite{halfaker2013rise} or alter the composition of editors working on particular content~\cite{shi2019wisdom}. 
Further, the rationale for protection is not always apparent to non-admins~\cite{das2016manipulation}, leading some seasoned editors to regard them as harmful~\cite{ppconsideredharm}.

\vspace{1mm}
\xhdr{Present work}
Given that page protections may be either beneficial (preventing disruptive editing) or harmful (preventing productive contributions), we analyze their usage and impact on the English Wikipedia.
Specifically, we ask:

\begin{itemize}
    \item[\textbf{RQ1:}] How are page protections enacted on Wikipedia?
    \item[\textbf{RQ2:}] What is the effect of page protections on article quality?
\end{itemize}

For this, we first obtain $299\,721$ edit protections on the English Wikipedia between October $2012$ and March $2023$ using a methodology introduced by Hill and Shaw~\cite{hill2015page}.
Second, we collect $127{,}098$ relevant user requests for page protection submitted during this period, along with their associated articles.
Third, we enhance every article revision with metadata, such as article quality~\cite{halfaker2020ores}, which enables us to create a time series of metrics for every article.

Using these diverse data sources, we first characterize the usage of page protections on the English Wikipedia (\textbf{RQ1}).
We then conduct a quasi-experimental study~\cite{hernan2016using} to estimate the causal effect of page protections on article quality (\textbf{RQ2)}.
We match articles that were protected after a user's request for page protection to similar articles with a declined request. 
Using a dynamic difference-in-differences design, we then estimate the effect of protection on article quality, based on a well-established automated quality metric that considers features about article structure and text~\cite{halfaker2020ores}.

\vspace{1mm}
\xhdr{Findings} 
First, analyzing page-protected articles (\textbf{RQ1}), we observe that protections are typically brief, often lasting for a week or less, with certain pages being repeatedly protected over the course of several years.
Edits generally increase before the enforcement of protections and decrease after the protection period concludes.
Second, through our difference-in-differences analysis, we discover that page protections' impact strongly depends on the article's nature in the pre-intervention period (\textbf{RQ2}).
Page protections are beneficial for high-quality articles, but detrimental to low-quality ones.
Subsequent analysis indicates that this could be because protections safeguard high-quality content (potentially by curbing vandalism) but prevent low-quality articles from improving (potentially by curbing contributions).
Furthermore, the effect varies across topics, with a more pronounced effect in articles related to ``Geography'' and ``History \& Society,'' a weaker effect in articles about ``Culture,'' and no notable effect on ``STEM'' articles.

\vspace{1mm}
\xhdr{Implications} 
Our research offers valuable insights for editors and administrators of Wikimedia projects and other interested researchers.
Specifically, our findings indicate that the quality and topic of an article mediate the effect of page protection on article quality, which could be explicitly considered when deciding whether to protect an article or not.
These findings can assist Wikipedia's stakeholders in making informed decisions and developing guidelines for page protection, promoting the continued enhancement of the largest encyclopedia on the Web.

\begin{figure}[t]
	\includegraphics[width=\linewidth]{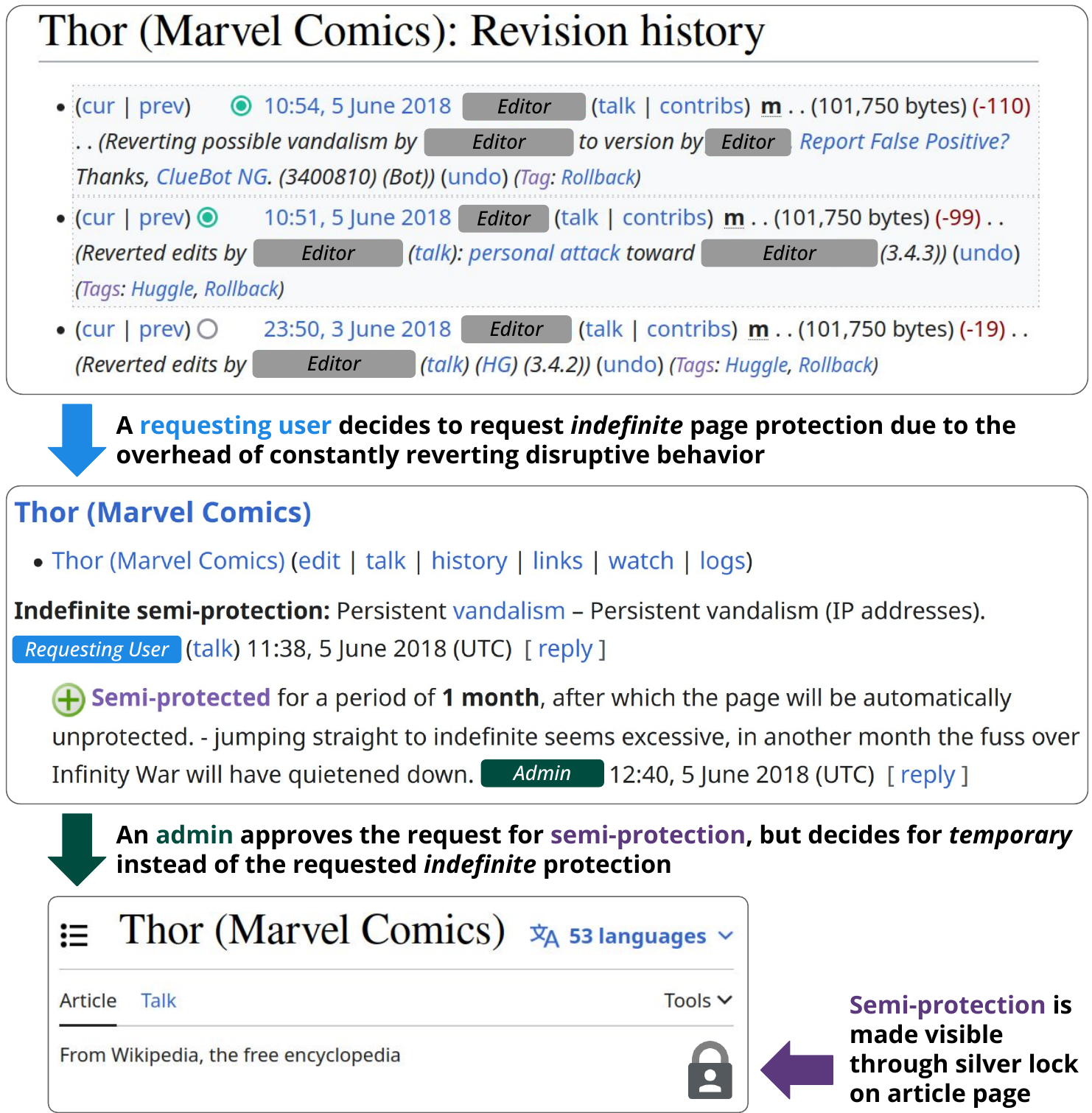}
	\caption{\textbf{Example of a request that led to a page being protected.} 
		In June 2018, the article ``Thor (Marvel Comics)'' was subject to disruptive editing because of increased attention,
        which regular editors (grey) had to clean up through reverts.
		Consequently, a user (blue) requested indefinite semi-protection to address the persistent vandalism by anonymous editors.
		Upon review, an admin (dark green) ruled against indefinite protection but still imposed a one-month semi-protection, which was highlighted on the article page through the ``silverlock''.
		Note that an admin could have also declined the request (i.e., not protecting the page) or could have taken alternative actions, such as blocking certain users.
		\\\footnotesize Screenshot from requests for page protection archive \hyperref[w.wiki/7EKd]{(https://w.wiki/7EKd})
  }
	\label{fig:rfpp_example}
\end{figure}

\begin{table*}[t]
	\caption{\textbf{Page protection levels on the English Wikipedia.} 
		For the most common levels of editing protection, the age of the user account and the number of edits determine whether an editor can make revisions to an article. 
		In the case of \emph{semi-protection} (\emph{extended confirmed protection}), editors must be \textit{confirmed} (\textit{extended confirmed}) by having an account for at least 4 days (30 days) and making at least 10 (500) edits.
		Pending changes protection is a special kind of extension of the Wikimedia software.
		This protection mode allows anyone to edit an article, but changes made by unregistered or unconfirmed editors must be approved by a pending changes reviewer before they become visible to readers who are not logged in.
		Higher protection levels (i.e., template, full, or interface) can only be bypassed by high-privilege user roles (template editor, admin, or interface admin). }
	\label{tab:en_pp_types}
	\resizebox{\textwidth}{!}{
		\begin{tabular}{lcccccc}
			\hline
			& \multicolumn{6}{c}{\textbf{User Type or Role}}      \\
			& \textbf{Unregistered/New} & \textbf{(Auto-)Confirmed} & \textbf{Extended confirmed} & \textbf{Template editor} & \textbf{Admin} & \textbf{Interface admin} \\ \hline
			\textbf{No protection}      & \cmark & \cmark & \cmark & \cmark & \cmark & \cmark \\
			\textbf{Pending changes}    & \imark & \cmark & \cmark & \cmark & \cmark & \cmark \\
			\textbf{Semi}               & \xmark & \cmark & \cmark & \cmark & \cmark & \cmark \\
			\textbf{Extended confirmed} & \xmark & \xmark & \cmark & \cmark & \cmark & \cmark \\
			\textbf{Template}           & \xmark & \xmark & \xmark & \cmark & \cmark & \cmark \\
			\textbf{Full}               & \xmark & \xmark & \xmark & \xmark & \cmark & \cmark \\
			\textbf{Interface}          & \xmark & \xmark & \xmark & \xmark & \xmark & \cmark \\ \hline
		\end{tabular}
		
	}
	\raggedright
    \xmark\ = Can not edit;
	\cmark\ = Can edit;
	\imark\ = Can edit, but changes are hidden from readers who are not logged in, until approved by a reviewer (logged in readers see changes right away). 
	Table adapted from \href{https://w.wiki/7QUz}{Wikipedia:Protection\_policy (https://w.wiki/7QUz)}.
\end{table*}
\newpage
\section{Page Protections and Requests}
\label{sec:pp}
We provide some background on the specifics of page protection and requests for page protection on the English Wikipedia.

\xhdr{Protection process}
Wikipedia is intended to be kept as open as possible~\cite{wikiPPPolicy}.
Thus, admins are generally urged to enforce protection only when absolutely necessary to prevent further damage.
Preemptive protection is generally not permitted (e.g., expected traffic due to current events), although certain articles are treated with greater caution~\cite{wikiPPPolicy}.
However, in cases of disruptive editing behavior by multiple editors, Wikipedia administrators may be compelled to implement some form of edit protection.\footnote{Other protections exist (e.g., move), but in this work we focus on edit protection.}
The most commonly enforced protection is \textit{semi-protection}, which requires an editor to have an account for at least four days and to have made at least ten edits.
Stricter levels of protection exist, such as \textit{full protection}, which only allows admins to edit content.

The level of protection enforced for an article depends on several factors, including the severity of the disruptive behavior, previous controversies, and past protections.
Table~\ref{tab:en_pp_types} illustrates the six protection levels in the English Wikipedia as of September 2023.
Admins enforce protection for either a temporary period, after which the article will be automatically unprotected, or for an indefinite period, which requires manual removal of the protection.
Additionally, admins can adjust the protection levels during active protection.
%

\xhdr{Requests for page protection}
While admins can enforce page protections independently, users can also submit requests for page protection (\rfpp). 
When an \rfpp\ is submitted through the request form~\cite{rfppForm}, it automatically creates an entry on the \rfpp\ overview page, which admins then review~\cite{rfppGeneral}.
After review, admins make judgments through responses elicited by coded templates, such as \textit{Declined}, \textit{Semi-protected} or \textit{Fully protected} (i.e., the request has been approved), \textit{User(s) blocked}, \textit{Already protected}, and others.\footnote{For a full list of possible responses, see: \url{https://w.wiki/6oRr}}
A single \rfpp\ can have multiple responses from the same or different admins.
For example, admins often seek additional comments and clarification from the requesting user or change their response within minutes of the initial decision.
Once the admin's decision resolves the request, Wikipedia archives the \rfpp\ on pages that are structurally similar to article talk pages~\cite{rfppArchive, viegas2007talk}.

Fig.~\ref{fig:rfpp_example} illustrates this procedure for an accepted \rfpp. 
In this case, the article ``Thor (Marvel Comics)'' experienced vandalism from multiple anonymous editors, resulting in a user submitting an \rfpp.
Although the admin rejected the request for indefinite semi-protection, they decided to impose a temporary one-month semi-protection instead.
Afterward, the ``silverlock'' signaling semi-protection appeared on the article page for ``Thor (Marvel Comics)''.

\xhdr{Protection in other language editions}
In this work, we focus on the English Wikipedia.
While page protections and requests for page protection exist in various language editions, differences in implementation may occur~\cite{Johnson2022}. 
For example, in German, only four protection types exist (instead of six in English), and users do not request protection but instead report user misconduct through a form~\cite{deutscheWikiMeldung}, making it difficult to compare multiple language editions.

\section{Related Work}


\xhdr{Wikipedia content policing}
Wikipedia uses a variety of policies and moderation tools to ensure productive user collaboration.
First, admins can permanently block users for severe or repeated offenses~\cite{das2016manipulation}.
Second, edit filters are mechanisms that detect whether contributions to a Wikipedia page violate certain criteria, such as whether they contain swear words or only use uppercase letters~\cite{vaseva2020you}.
Depending on their configuration, the user will receive a warning, the edit summary of the revision will be flagged in the article revision, or the revision will be denied altogether.
Third, many language versions use bots to automatically police content~\cite{zheng2019roles}.
For example, \mbox{\emph{ClueBotNG}}---one of the most notable bots in the English Wikipedia---automatically reverts vandalism~\cite{cluebotNG}.
Certain tools (such as bots), as well as researchers, use the machine learning framework \textit{ORES} to evaluate Wikipedia content and editor activity~\cite{halfaker2020ores}.
ORES provides endpoints for several machine learning models that can be used to predict both article characteristics (e.g., topic or quality) and edit characteristics (e.g., whether it is damaging or made in good faith).
In general, while admins attempt to sanction rogue users and vandalism swiftly, controversies can arise from valid changes, and their reversal can lead to edit wars~\cite{yasseri2013value}. 

\xhdr{Effects of page protections on Wikipedia}
Early Wikipedia studies theorized rising page protections as a possible indication of increasing administrative action~\cite{suh2009singularity}. 
Furthermore, page protection can present additional obstacles for newcomers who already have difficulty integrating into the editorial community~\cite{halfaker2013rise}.
Hill and Shaw \cite{hill2015page} describe the relevance of page protections and propose an approach to collect accurate accounts of protection ``spells'' for the English Wikipedia. 
They find that while only about $0.67\%$ of articles on the English Wikipedia were protected in December 2013, these articles accounted for about $14.3\%$ of all views on the online encyclopedia.
Most relevant to our work, a recent study used a combination of qualitative and quantitative methods to assess the impact of protections in a single article category (``Internet Culture'')~\cite{ajmani2023peer}.
The authors find that for this specific subset of articles, protections lead to high editor turnover and protection requests can be categorized according to editor activity, article topic, and article visibility.
Overall, the literature to date has proposed data collection approaches and conducted parsimonious analyses of the impact of page protections on editors.

\xhdr{Predicting page protections}
The process of manually enforcing page protection is tedious for both requesting users and admins.
Thus, Spezzano et al. \cite{spezzano2019detecting} built a machine learning classifier to predict whether a page will (or should) be protected.
Their classifiers suggest that the most important features for classifying articles as worthy of protection are the rapidity of contributions (mean time between revisions), anonymous edits, and article topic.

\xhdr{Protection as a proxy of managerial authority}
DeDeo leverages page protections as an example of top-down authority, which is driven by reverts \cite{dedeo2016conflict}.
His analysis (62 articles) suggests that both the start and end of protections rarely have lasting effects on long-term activity. 
Similarly, Klapper and Reizig view page protection as enforcement of lateral authority~\cite{klapper2018effects} and find that affected editors decrease their activity on talk pages of protected articles but increase their activity elsewhere.

\xhdr{Restrictions in other Web platforms}
Other community-driven websites employ mechanisms comparable to page protections.
On the discussion platform Reddit, admins possess the ability to restrict (i.e., only selected users can create new posts), quarantine (i.e., hide on Reddit's search and front page), or completely ban sub-forums.
This has been associated with a decline in activity and migration of users to alternative platforms~\cite{horta2021platform, chandrasekharan2017you}.
In contrast, automated content moderation on Facebook has been shown to promote adherence to rules without decreasing commenting activity~\cite{ribeiro2022automated}.

\begin{table}[b]
	\caption{\textbf{Datasets.} 
    We provide an outline of the full dataset of protections and requests (Section~\ref{sec:data}) as well as the matched dataset for our quasi-experimental study (Section~\ref{sec:results_all}).
    }
	\label{tab:dataset_stats}
\begin{tabular}{l|rr}
\hline
\textbf{\begin{tabular}[c]{@{}l@{}}Edit \\ Protections\end{tabular}} & \multicolumn{1}{c}{\textbf{\begin{tabular}[c]{@{}c@{}}Temporary\\ (Ended)\end{tabular}}} & \multicolumn{1}{c}{\textbf{\begin{tabular}[c]{@{}c@{}}Matched\\ Study\end{tabular}}} \\ \hline
Obs. period & 01/2012 – 03/2023 & 01/2013 – 12/2022 \\
Count & 196{,}240 & 102{,}156 \\ \hline
\textbf{\begin{tabular}[c]{@{}l@{}}Requests for\\ Page Protection\end{tabular}} & \multicolumn{1}{c}{\textbf{\begin{tabular}[c]{@{}c@{}}Parsed\\ from Archives\end{tabular}}} & \multicolumn{1}{c}{\textbf{\begin{tabular}[c]{@{}c@{}}Matched\\ Study\end{tabular}}} \\ \hline
Obs. period & 08/2012 – 03/2023 & 01/2013 – 12/2022 \\
Count & $127{,}098$ & $48{,}308$ \\
\ \ \ \ \% Accepted & $65.69$ & $50.00$ \\
\ \ \ \ \% Declined & $20.23$ & $50.00$ \\
\ \ \ \ \% Others & $14.08$ & 00.00 \\ \hline
\end{tabular}
\end{table}

\xhdr{Contributions of our work}
We extend these previous works by (i) providing code to retrieve and parse page protection requests that can be linked to the page protection dataset by Hill and Shaw~\cite{hill2015page} alongside ORES scoring of revisions, (ii) describing the characteristics of page protections and requests from January 2012 to March 2023, and (iii) conducting a quasi-experimental study to provide the first analysis of how page protections affect article quality.

\section{Characterizing Page Protections}
\label{sec:data}

\begin{figure}[t]
	\includegraphics[width=\linewidth,clip,trim={0 0 4cm 0},]{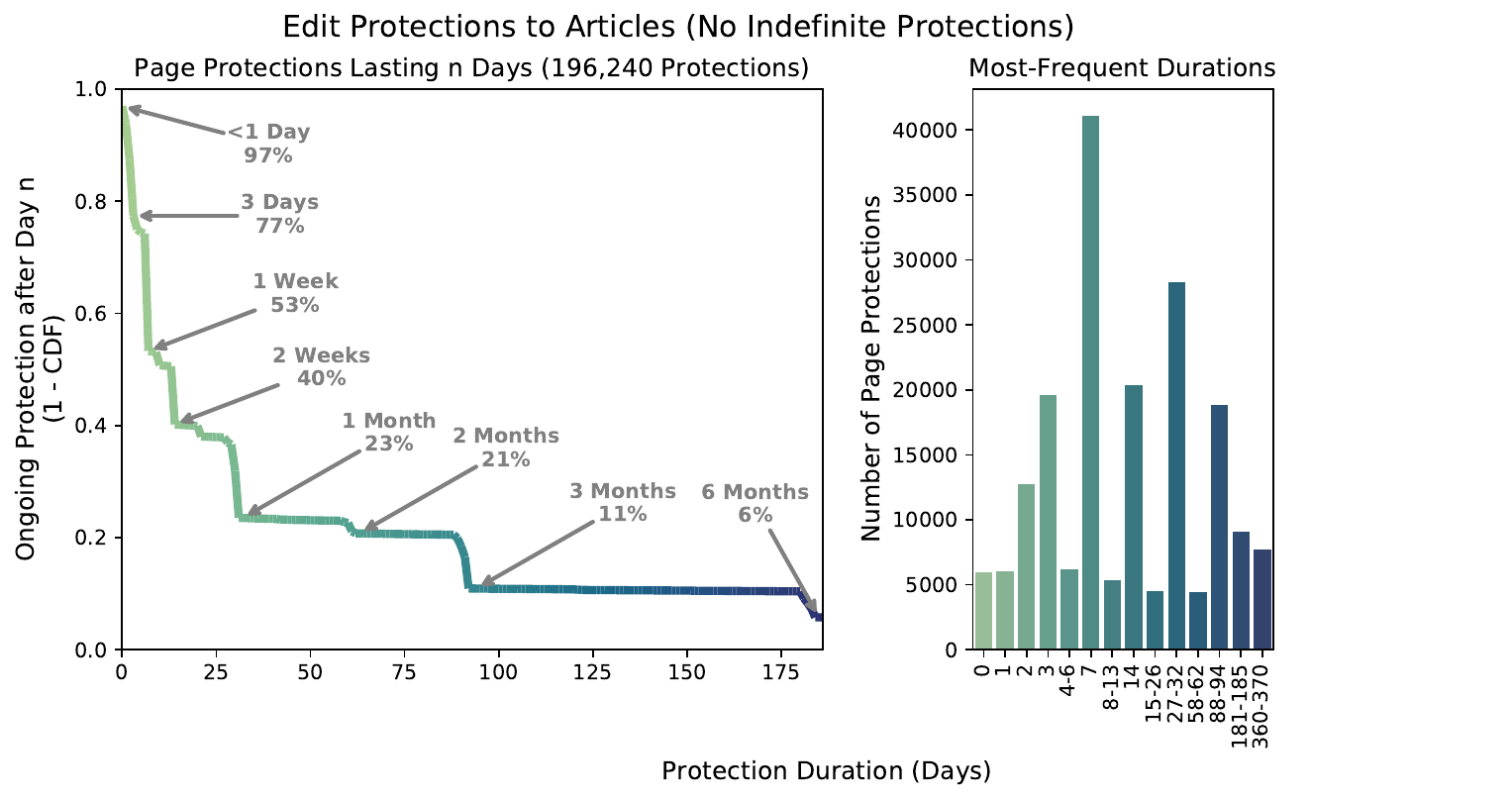}
	\caption{\textbf{Edit protections on the English Wikipedia.}
		We show the duration distribution of $\mathbf{196{,}240}$ temporary article protections between January 2012 and March 2023. 
        Most protections are enforced for a week, but three-day, two-week, one-month, and three-month protections also occur regularly.} 
	\label{fig:pp_data_overview}
\end{figure}
We now characterize page protections while discussing our dataset and preprocessing.
Our results are based on the openly available Wikipedia dumps, and all the resources (data, descriptive statistics, and code) required to reproduce the analyses in this paper are available at \url{https://github.com/ruptho/wiki-pp}. 
To avoid misuse of the dataset in terms of user anonymity, we do not publish any parsed usernames.

\xhdr{Page protection spells}
We process page protection spells similarly to Hill and Shaw~\cite{hill2015page}.
This approach only includes protections with at least the level of semi-protection and leaves out pending changes protections.
We obtain $488{,}772$ protection spells starting after December 31\textsuperscript{st} 2011, and for this work, only consider edit protections ($299{,}721$), dropping page move or creation protections.
Of these $299{,}721$ spells, $215{,}746$ regard article pages (``namespace 0''), while the remainder concern other pages, such as images or talk pages.
We further remove spells for articles that were moved (i.e., name changes) during a spell (total of $2{,}502$ protections).
Of the remaining spells, $17{,}004$ protections are active as of March 2023, with a median protection time of $1{,}003$ days and $932$ spells being upheld for less than $32$ days. 
We exclude these indefinite protections in our final dataset of $196{,}240$ protections (Fig.~\ref{fig:pp_data_overview}; Table~\ref{tab:dataset_stats}).

These $196{,}240$ spells were applied to a total of $104{,}305$ unique pages---with the articles about Basketball player \textit{DeMarcus Cousins} and the country \textit{Turkey} being the most frequently protected ($30$ and $29$ spells, resp.).\footnote{DeMarcus Cousins: \url{https://w.wiki/3ovx}, Turkey: \url{https://w.wiki/3hLW}}
Meanwhile, $67{,}131$ articles were only protected once.
Overall, this signals a frequent rotation of temporary protections while certain articles have longer, indefinite protection.
This aligns with Wikipedia guidelines, as they consider exuberant, long, or indefinite protections of pages harmful and state that such protection patterns should only be applied where warranted (e.g., the main page) or in case of repetitive disruptive behavior even after temporal protections expire~\cite{ppconsideredharm, wikiPPPolicy}.

\xhdr{Page protection requests}
We parse $163{,}227$ requests for page protection (\rfpp) from the English Wikipedia archives~\cite{rfppArchive} from the beginning of their recording (October 2012) until March 2023.
As the archives are structured similarly to Wikipedia talk pages, we adopt and extend existing code from MWChatter\footnote{\url{https://github.com/mediawiki-utilities/python-mwchatter}} and mwparserfromhell\footnote{\url{https://github.com/earwig/mwparserfromhell}} to parse their wikitext (i.e., Wikipedia's ``source code'').
We discard requests  
(1)~with malformed wikitext or article names, without a valid timestamp for the users' initial request, or without a final decision by an admin ($3.7\%$ of archived requests);
(2)~for pages not in the article namespace ($21{,}520$);
(3)~from actions other than editing ($6{,}815$); and
(4)~for pending changes protection ($7{,}794$).
We then select the latest admin response to the resulting $127{,}098$ valid \rfpp s and group them into five groups~(Appendix~\ref{app:data}, Fig.~\ref{fig:pp_rfpp_overview}; Table~\ref{tab:dataset_stats}): protected ($83{,}487$), declined ($25{,}718$), declined but carried out an intervention affecting certain users instead (e.g., user block or referral to the edit-warring noticeboard; $8{,}169$), protection existing ($8{,}096$), or others (e.g., withdrawal of request; $1{,}628$). 

\begin{figure}[t]
\centering
	\begin{subfigure}[b]{.49\columnwidth}
		\centering
		\includegraphics[width=\textwidth]{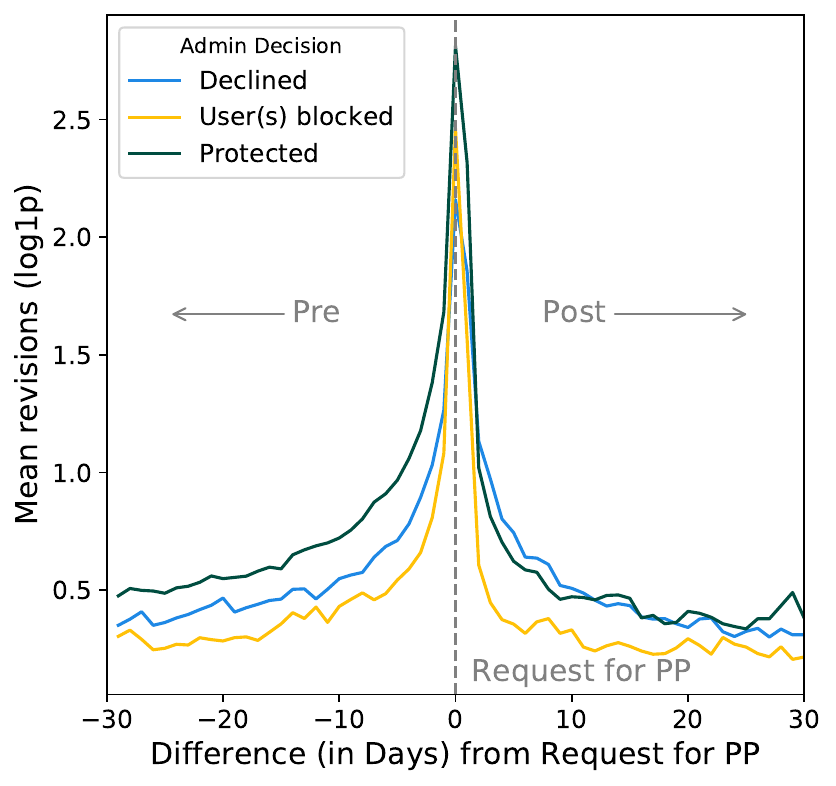}
		\caption{Requests for page protection.}
		\label{fig:pp_rfpp}
	\end{subfigure}
	\begin{subfigure}[b]{.49\columnwidth}
		\centering
		\includegraphics[width=\textwidth]{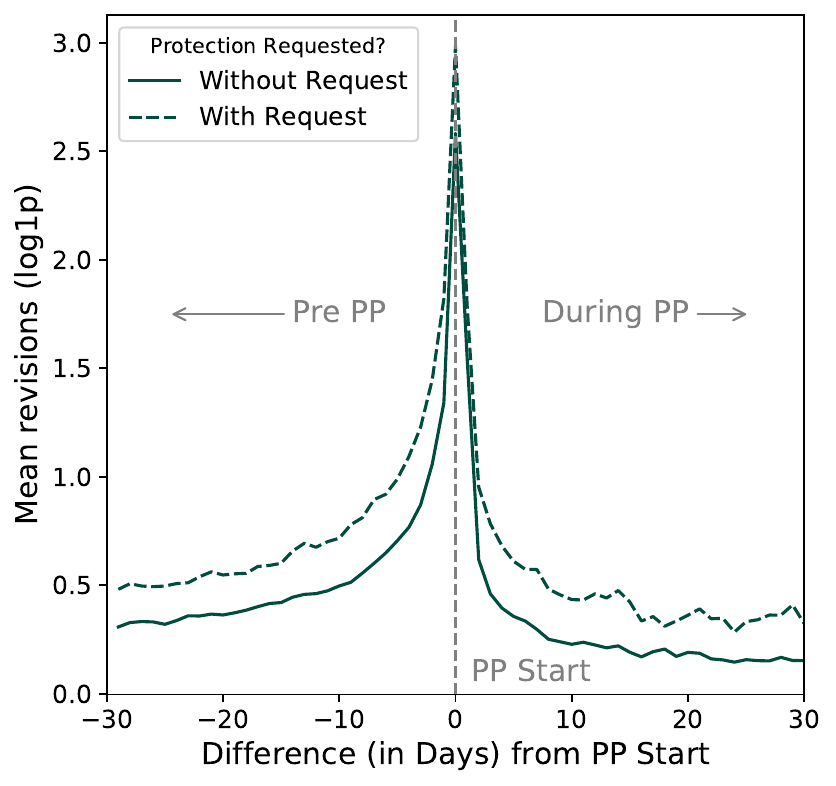}
		\caption{Enforced page protections.}
		\label{fig:pp_spells}
	\end{subfigure}
	\caption{\textbf{Edits around page protection key events.}
		We observe that both for \rfpp\ \emph{(a)} as well as enforced protections \emph{(b)}, spikes in activity are dampened by the corresponding event, and edits revert to levels prior to the intervention.
  }
	\label{fig:pp_overview_revisions}
\end{figure}

\xhdr{Merging requests and first protections}
To exclude confounders that might arise due to multiple protections such as previous history of disruptive editing, we only include the first semi-protections of articles ($94{,}298$ spells) in our analysis of the effect of page protections on quality.
We thus merge first protection spells that started after the implementation of the \rfpp\ form in October 2012 ($88{,}239$ spells) with accepted requests for page protections.
We find requests for $34{,}401$ spells ($39\%$ of first protections), suggesting that most first protections are actually enforced by admins rather than requested by users.
While we are unable to match around $4\%$ of accepted \rfpp\ to spells, and we previously noted that we could not parse around $3\%$ of the original \rfpp s from the archives, this still represents a considerable amount of non-requested page protections.

\xhdr{Article metadata} 
We utilize the Mediawiki history dataset~\cite{mwdumps} as well as the ORES API~\cite{halfaker2020ores} to collect metadata for our articles~.
We use the dumps to gather revision data such as edit counts and article length and leverage ORES to predict article quality and topic.

We collect the predicted top-level topic labels based on ORES' \emph{articletopic} taxonomy (\textit{Culture}, \textit{STEM}, \textit{History and Society}, and \textit{Geography}).\footnote{\href{https://w.wiki/7Rrd}{ORES Articletopic endpoint (https://w.wiki/7Rrd)}}
We only assign a topic to a revision if ORES predicts a likelihood for the topic greater than 0.5 for the majority of five revisions made before the request or start of protection, and drop all articles that without a topic.
For article quality, we use ORES' \emph{articlequality} model, which predicts quality from structural features of articles (e.g., number of sections and references or article length) rather than actual writing or factual accuracy.
In practice, however, these metrics correlate with nuanced aspects of the texts~\cite{wikiORES}.
The \emph{articlequality} endpoint predicts a label based on the probability assigned to one of six quality classes present in the English Wikipedia: Start, Stub, C-, B-, Good, and Featured Articles (in ascending order).
As research has demonstrated that predicting categories may be prone to mislabeling~\cite{warncke2013tell,halfaker2017interpolating}, we follow the approach by Halfaker to combine category probabilities into a single quality score between 0 and~5~\cite{halfaker2017interpolating}.

\xhdr{Editing patterns by admin decision and protection type}
We analyze editing activity by notable admin decisions for first protections around the \rfpp\ and the start of protection~(Fig.~\ref{fig:pp_overview_revisions}). 
In particular, Fig.~\ref{fig:pp_rfpp} suggests that \rfpp s are preceded by edit spikes, followed by a decrease immediately after the request, regardless of an admin's decision.
Second, there appear to be lower activity levels in articles that are protected without a prior \rfpp\ compared to protections of articles with an \rfpp\ (Fig.~\ref{fig:pp_spells}).
This suggests that, regardless of administrative decision, \rfpp\ affects activity. 

\section{Protections and Article Quality}
\label{sec:results_all}

\begin{figure}[t]
	\includegraphics[width=.5075\columnwidth, trim={.2cm 0 .2cm 0cm},clip, ]{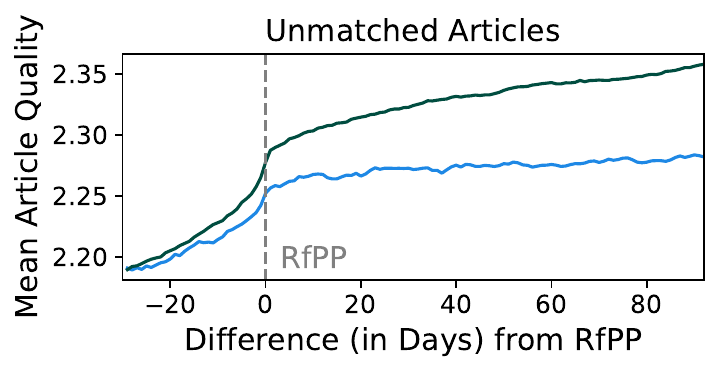}\hfill
    \includegraphics[width=.4825\columnwidth, trim={.9cm 0 .2cm 0cm},clip, ]{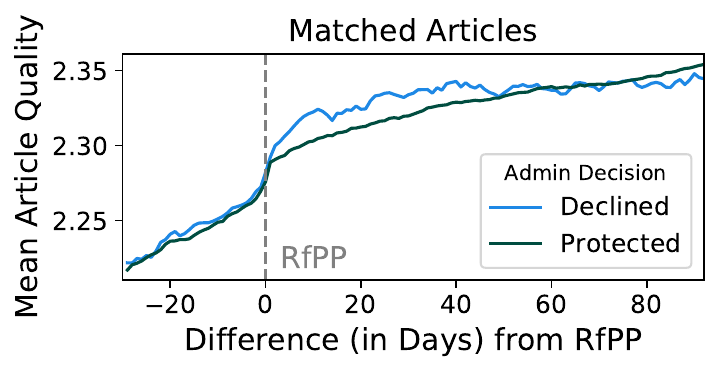}
 
	\caption{
    \textbf{Unmatched and matched mean article quality}.
	We note differences in trends in the unmatched data before the request for page protection (\rfpp), which are not evident in the matched data, indicating good quality of the match.
    \label{fig:matching}
    }
\end{figure}

Next, we describe our quasi-experimental design and present the results of our difference\hyp{in\hyp{differences}} analysis on the effect of page protections on article quality.
Finally, we discuss potential interpretations of our results and suggest future research, while also outlining possible limitations of our study.


\begin{figure*}[t] 
	\centering
	\begin{subfigure}[t]{.325\textwidth}
		\includegraphics[trim={0 0 0 .65cm},clip, width=\textwidth]{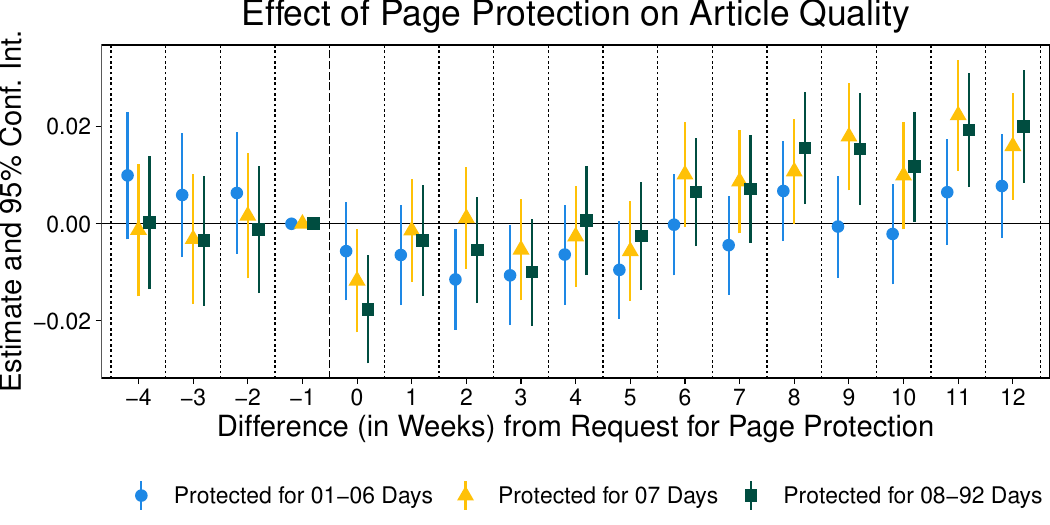}
		\subcaption{Effects for All Articles.}
			\label{fig:effect_aq_all}
	\end{subfigure}\hfill
	\begin{subfigure}[t]{.325\textwidth}
		\includegraphics[trim={0 0 0 .65cm},clip,width=\textwidth]{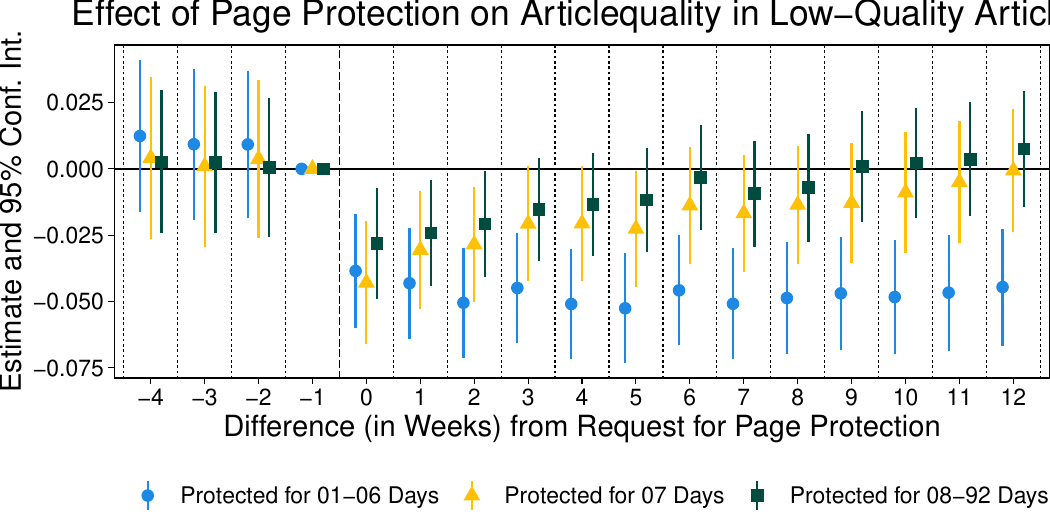}
		\subcaption{Effects for Low-Quality Articles.}
        \label{fig:effect_aq_low}
	\end{subfigure}\hfill
	\begin{subfigure}[t]{.325\textwidth}
		\includegraphics[trim={0 0 0 .65cm},clip,width=\textwidth]{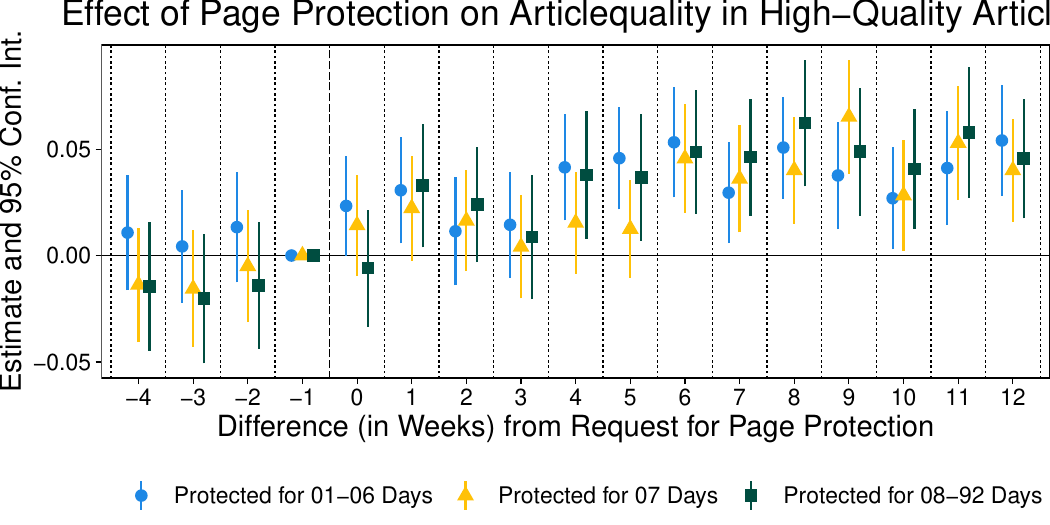}
 	\subcaption{Effects for High-Quality Articles.}
        \label{fig:effect_aq_high}
	\end{subfigure}
	\caption{\textbf{Difference in article quality after protection.}
		We use difference-in-differences models to analyze effects on the matched dataset and in (a) find small differences in quality between articles with declined and accepted requests. 
		In (b) and (c), we fit the same model on low-quality and high-quality articles, and detect heterogeneous effects.
	}
	\label{fig:effect_aq}
\end{figure*}
\subsection{Quasi-Experimental Setup}
\label{sec:experiment}

We create a quasi-experimental study (see Hernán~\cite{hernan2016using}) that simulates the page protection process as the following randomized experiment:
when admins receive a \rfpp, they flip a coin and accept the \rfpp\ if it lands on ``heads'' or reject the \rfpp\ if it lands on ``tails.'' 
Then, in the 13-week period following the decision, they track the article quality of protected and non-protected pages to determine the effect of protection on quality.
Note that this experiment is not practically feasible, as \rfpp s are not accepted or declined ``at random.''
Therefore, we describe how we approximate this experiment using matching and difference-in-differences regression.

\xhdr{Data}
We select the observational period from January 2013 until December 2022, as the English Wikipedia implemented \rfpp\ in late 2012, and the processed dumps extend until March 2023.
We analyze articles that were not previously protected and disregard \rfpp\ of articles that were protected or received another \rfpp\ within 90 days to minimize potential confounding effects from prior protection phases or special administrative oversight.
Additionally, we exclude first protections longer than 91 days.
Our experiment's pre-treatment period spanned four weeks before up until the actual request, and we cover 13 weeks (91 days) in the follow-up period after the \rfpp, where we consider an article to be consistently treated.
Early investigations suggested that the end of protection does not result in significant changes in activity or article quality trends (see Appendix~\ref{app:data}).
We consider the \rfpp\ as the time of treatment, as administrators can initiate clean-up activities immediately after a protection request is submitted, even before protection is formally applied.
Nonetheless, protections are generally enforced close to the \rfpp\ (median of $117$ minutes).
Finally, the target variable of our study is the last recorded quality for an article per week.

\xhdr{Matching}
The data described above allows us to compare articles that had their protection requests approved with those that had not. 
However, a simple comparison may be problematic, as unlike in our hypothetical randomized experiment, in reality page protections are not assigned randomly.
On the contrary, there may be confounders that can influence both the assignment of treatment and outcome. 
To account for these, we use the pre-treatment period to match protected (``treated'') to non-protected articles (``control'') using the following features measured before the \rfpp:
\begin{itemize}[leftmargin=*, itemsep=0em]
	\item \textbf{Activity.}
    Edits 1 hour, 24 hours, and 1 week before ($\log(x+1)$).
	\item \textbf{Controversy.}
    Number of identity reverts 1 hour, 24 hours, and 1 week before, as it is considered a proxy of conflict~\cite{sumi2011edit} ($\log$).
	\item \textbf{Article length.}
    Maximum article length in bytes ($\log$).
	\item \textbf{Article age.}
    Overall number of edits to the article ($\log$).
	\item \textbf{Quality.} 
    Maximum article quality 1, 8, and 21 days before.
	\item \textbf{Topic.}
    Top-level topics (History and Society, STEM, Culture, Geography). An article can have more than one topic.
\end{itemize} 
We employ propensity score matching~\cite{hill2006interval, stuart2011matchit} but require exact matching on article topic(s) and use caliper matching for quality (caliper $=0.25$, one-quarter of a quality level) as well as the propensity score (one standard deviation).
After matching, all standardized mean differences for the covariates were below 0.1~\cite{stuart2011matchit}.
This approach produces $24{,}154$ matched control-treatment pairs after discarding $1{,}401$ rejected (control) and $1{,}463$ accepted (treatment) \rfpp s for which we did not find matches~(Table~\ref{tab:dataset_stats}).
We plot the mean article quality for matched and unmatched articles in Fig.~\ref{fig:matching}.

\xhdr{Difference-in-differences model}
Comparing outcomes between treatment and control groups could uncover the effect of page protections, provided that the matching variables control for all causal paths between treatment and outcome~(\cite{pearl2009causality}; Appendix~\ref{app:methods}).
Nonetheless, we use a difference-in-differences approach to make our results more robust.
We compare treated articles (protected) with control articles (not protected) under the parallel trends assumption, i.e., that in the absence of treatment, the difference between the treatment and control group remains constant over time.
Intuitively, this method creates a counterfactual estimate of how the treatment group would have progressed without treatment and compares it to the observed change in the outcome. 

In practice, the difference-in-differences approach enables us to estimate the causal impact of the protection on the article quality using a simple linear model. We consider a panel of $a=1,\ldots, N$ units for $t\in T$ relative periods and estimate coefficients $ \delta_{t}$ using a fixed effect regression of the form
\begin{equation}
    \label{eq:dd2}
    Y_{a,t} = \alpha_a  + 
    \sum_{t \in T} \delta_{t}  D_{t} + 
    \epsilon_{a,t},
\end{equation}
where, $Y_{a,t}$ is the outcome associated with article $a$ at time $t$, $\alpha_a$ are unit fixed effects, and
$D_{t}$ are indicator variables for unit $a$ being $t$ periods before or after treatment.
We measure the effect of treatment on outcome in time $t$ via $\delta_{t}$ and use the R package \emph{fixest}~\cite{fixest} for computation.
Due to our matching, we assume that the parallel trends condition holds before the intervention, and we attempt to prevent anticipation for the treatment (protection) by setting our treatment to the time of the \rfpp\ instead of the protection start.
We do not account for time-fixed effects as we consider the relative date of the request for page protection.

\subsection{Results}
\label{sec:results}
We now present the measured effects of page protection on article quality. 
Note that we report results for z-score normalized values of our target variable (i.e., quality), meaning that an increase of $0.1$ equals an increase by 10\% of the standard deviation.

\xhdr{Short-term decrease, long-term recovery}
In Fig.~\ref{fig:effect_aq_all}, we visualize weekly effects for all protections lasting $<7$ days ($8{,}949$ pairs), $7$ days ($8{,}401$ pairs), and $> 7$ days ($6{,}804$ pairs).
We observe that for protections lasting $7$ or $>7$ days, quality shows a minor but significant decrease in the week after the request ($-0.017$ and $-0.018$, $p<0.05$, resp.), while protections of less than $7$ days observe no significant changes ($-0.006$, $p>0.05$).
In the weeks thereafter, as quality recovers to pre-request baselines around week $6$, articles subject to protections for $7$ days or longer even significantly raise their quality until week $12$ ($0.016$ and $0.02$, $p<0.05$, resp.).
Overall, we conclude that page protections in the short term seem to reduce quality, but in the long term, to increase quality. 
However, although reported effects are significant, the overall effect size is small (e.g., an increase of around 2\% of the standard deviation in week 12).

\xhdr{Low-quality articles suffer, high-quality articles improve}
We now categorize articles into low quality (lowest quartile, quality $<1.71$; Fig.~\ref{fig:effect_aq_low}) and high quality (highest quartile, quality $> 2.94$; Fig.~\ref{fig:effect_aq_high}).
Low-quality articles experience slight but significant decreases right after protection (e.g., $-0.043, p<0.05$ for 7-day protections).
By week 12, the levels recover to baseline for most, except for those with less than 7 protection days, which sustain a level of quality similar to the one observed during the initial decline ($-0.045, p<0.05$ in week 12).
For high-quality articles, we found no significant changes in the week after the protection ($p>0.05$ for all durations). 
Instead, these articles slightly improve their long-term quality (e.g., 7-day protections in week 12, $0.058, p>0.05$).

\begin{figure}[t]
	\includegraphics[width=\columnwidth]{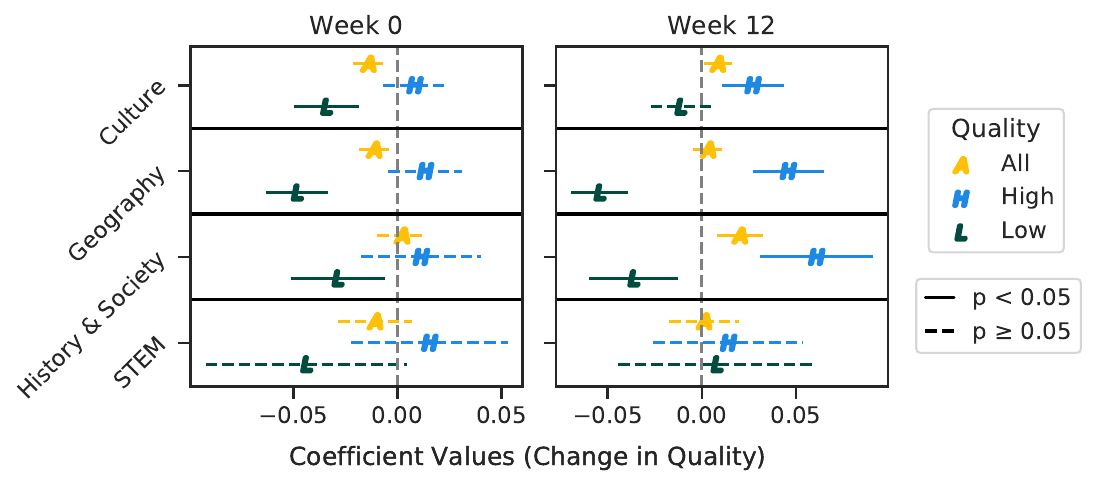}
	 \caption{Effect of page protection by topic.
           Low-quality articles show a decrease right after protection (week 0) and mostly do not improve in the long-term (week 12), while high-quality articles generally improve.
           Notably, STEM is an exception in both subsets, as it remains relatively stable.
           }
	\label{fig:effect_topics}
\end{figure}

\xhdr{Differential effects for topics}
Topics strongly shape attention and content on Wikipedia~\cite{kobayashi2021modeling, singer2017we}.
Therefore, we combine all protection durations and investigate effects for Culture, Geography, History \& Society, and STEM for all articles as well as only low-quality and high-quality articles~(Fig.~\ref{fig:effect_topics}).
We find mostly homogeneous effects for topics in low-quality articles right after protection, with small decreases and a more negative effect in the long term (e.g., Geography $-0.048$ for week 0, $-0.062$ for week 12).
For various topics in high-quality articles, we observe improved quality over time, with most topics showing a gradual increase in quality (e.g., History and Society, $0.061, p>0.05$ for week 12).
Notably, effects are not significant for STEM articles ($p>0.05$).

\xhdr{Article length, other factors also influence quality}
Next, we examine an important component of article quality: length. 
Similar to our previous approach, we categorize articles into quartiles according to pre-\rfpp\ length ($\log$), dividing them into shorter (lowest quartile, $<9.96$) and longer articles (highest quartile, $>10.34$), and employ our previous model with length as the target variable ($\log$ scale, z-score standardization; Appendix~\ref{app:results}, Fig.~\ref{fig:effect_size}).
Overall, the results are similar to what we found for quality.
Articles that were already long prior to protection became even lengthier by week 12 ($0.07, p<0.05$; Fig.~\ref{fig:effect_size}), whereas short articles did not grow. 

However, we find that length does not fully explain the increased quality we observe.
For all 26 article features employed by ORES' \emph{articlequality} model, such as the number of sections, links, and references~\cite{wikiORES, halfaker2020ores}, we perform a variation of our difference\hyp{in\hyp{difference}} analysis, comparing \emph{week -1} with \emph{week 0} and \emph{week 12}, respectively (features are $\log$-scale, z-score standardized).
We find that our overall quality is also driven by other features such as the number of references and links~(Appendix~\ref{app:results}, Fig.~\ref{fig:aq_all_coefs}).
For example, after protection, low-quality articles contain more ``citation needed' templates~\cite{redi2019citation} and have fewer references (week 0, $p>0.05$).
On the contrary, in the weeks following protection, high-quality articles significantly increase the number of headings, references, and links, among other things (week 12, $p<0.05$).

\xhdr{Robustness of the analyses}
We varied our matching configuration and difference\hyp{in\hyp{differences}} model to ensure robustness.
First, we placed narrower ($0.1$) and wider ($0.5$) calipers on the quality prior to the request.
Next, we performed exact matching on the year and month of the request.
Additionally, we fit a daily version of the model along with the weekly one.
We did not observe any considerable changes to our main findings in any of the variations.

\subsection{Discussion}
We now discuss our findings and their implications, examine some of our study's limitations, and highlight future avenues for research.

\begin{figure}[t]
	\includegraphics[width=.5075\columnwidth, trim={.2cm 0 .2cm 0cm},clip, ]{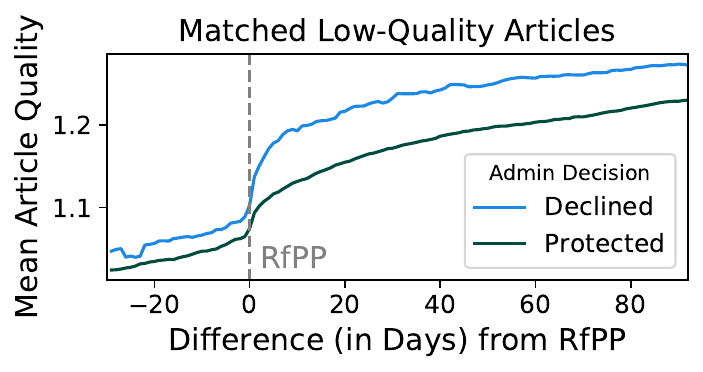}\hfill
    \includegraphics[width=.4825\columnwidth, trim={.9cm 0 .2cm 0cm},clip, ]{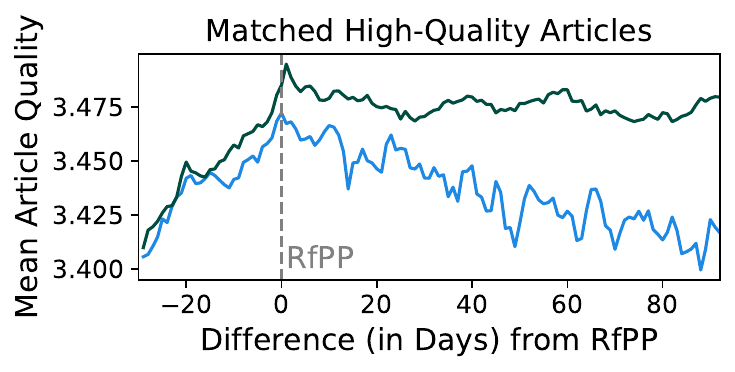}
 
	\caption{
    \textbf{Article quality in low- and high-quality articles by protection status}.
    Right after their request for page protection (\rfpp), low-quality articles display a slightly greater improvement in mean article quality compared to their unprotected counterparts (left), while unprotected high-quality articles undergo a decline in quality after the \rfpp\ and protected ones maintain their quality (right).
 }
 \label{fig:matching-high-low}
\end{figure}

\xhdr{Why low- and high-quality articles differ in effects}
We find that page protections have a positive effect on high-quality articles but a negative effect on low-quality ones. 
To better understand the mechanism driving this differential effect, we further analyze the mean article quality in our matched dataset, splitting by quality level and protection status (Fig.~\ref{fig:matching-high-low}).
This secondary analysis reveals an important observation: protected high-quality articles display resilience in maintaining their quality levels after the intervention, while non-protected high-quality articles exhibit a consistent decline in quality.
Conversely, trends for low-quality articles indicate that unprotected articles experience a bigger increase in quality after protection, compared to protected articles.

These patterns could be due to low-quality articles having greater potential for improvement when unprotected, as even small changes can be valuable to the content.
In that context, page protection reduces these articles' already limited chance of improvement, as on Wikipedia, a vast mass of articles are competing for attention at any time~\cite{ribeiro2021sudden}.
Furthermore, possible vandalism reverts or misinformation corrections by experienced editors or admins after protection could result in collateral damage due to excessive content removal.
This issue may affect shorter, low-quality articles more heavily than longer, high-quality ones. 
While high-quality articles can usually withstand the removal of a sentence or two, shorter ones may be disproportionately impacted.
In contrast, protected high-quality articles fare better than their unprotected counterparts, as without protection, new contributions may be of lower quality or vandalize the text (e.g., fewer references and links).
For these better articles, protection can even be viewed as a structured approach for quality improvement~\cite{warncke2015success}, which, for example, is triggered when articles compete for promotion to the featured article of the day~\cite{faToday}.

\xhdr{Not all topics are volatile}
The development of protected articles relating to History and Society, Culture, or Geography also exhibits this differential effect.
Across these topics, protections appear to have a detrimental impact on low-quality articles, but a positive effect on high-quality articles.
In contrast, STEM articles seem less prone to quality changes.
This could be attributed to the fact that STEM articles usually comprise more universal truths based on the nature of their subject matter, resulting in more consistent content. 
Conversely, topics such as ``Culture'' or ``History \& Society'' frequently involve articles about subjects susceptible to change and controversy, such as biographies of living persons and articles about wars or conflicts.

\xhdr{Ramifications for editors}
Besides directly affecting content, page protection's administrative barriers can hamper the overall growth of Wikipedia by discouraging editors, as previous work has shown that a similar negative impact on engagement occurs when edits by new users are reverted~\cite{halfaker2013rise}.
This is not typically a concern for high-quality articles, as they often cover popular subjects or receive frequent surges in attention that draw new users and content.
However, it may pose a serious problem for niche topics that receive little contribution.
Although past research has suggested that protection does not necessarily drive away editors~\cite{ajmani2023peer, klapper2018effects}, its effects may vary between novice and experienced users, analogous to the variations observed in low- and high-quality articles.
Further examining the effect of protections on the editor population is a promising avenue for future work.

\xhdr{Admin decisions and their motivation}
Wikipedia admins act as judges in the case of article protection.
As studies have shown the influence of biases and ideology on the decisions made by judges in a court of law~\cite{harris2019bias}, such factors may also affect the actions of Wikipedia admins~\cite{das2016manipulation} and therefore merit further investigation.

Moreover, in Section~\ref{sec:data}, we were only able to find \rfpp s for less than half of our edit protections.
This observation is significant, as it suggests reduced community participation in the protection process and frequent admin decision-making without editor input.
Protection of articles without a previous \rfpp s could result from spillover activity~\cite{zhu2020content} from other protected articles, for which a \rfpp\ may or may not have been submitted. 
Alternatively, administrators may protect these articles based solely on their own judgments~\cite{das2016manipulation} or after being notified of conflicts through ``watching'' or ``patrolling'' certain articles.\footnote{\hyperlink{https://w.wiki/4W33}{Help:Watchlist (https://w.wiki/4W33)}, \hyperlink{https://w.wiki/7kDy}{Wikipedia:Patrols (https://w.wiki/7kDy)}}
In combination with the findings of this study, future work should aim at uncovering these particular protections.

\xhdr{Practical implications}
Our research has practical implications for page protection guidelines on Wikipedia.
In addition to the existing policies~\cite{wikiPPPolicy}, admins should consider an article's characteristics before deciding on protection.
The findings of our study indicate that protecting high-quality articles carries minimal risk. 
However, greater caution seems necessary when protecting low-quality articles, particularly on specific topics.

\xhdr{Broader perspective and ethics}
By enhancing knowledge on page protections, this study contributes to Wikipedia's ongoing efforts to understand and assess knowledge integrity~\cite{aragon2021preliminary}. 
Concerning ethical or privacy concerns, we only employ data provided through Wikimedia APIs or dumps, and anonymize user information.

\xhdr{Limitations}
First, our primary measure is the well-established ORES article quality metric~\cite{teblunthuis2021measuring, halfaker2020ores, halfaker2017interpolating, warncke2013tell}.
However, this metric utilizes structural content features and does not assess knowledge integrity~\cite{aragon2021preliminary}, such as writing quality or factual correctness.
Nonetheless, in practice, basic structural features are associated with more advanced article characteristics~\cite{wikiORES, redi2019citation}.
This is also supported by the ORES model, as even though it does not consider the factual accuracy of an article, it does take into account the quantity of references and citations~\cite{piccardi2020quantifying} as well as the templates identifying missing references~\cite{redi2019citation}. 
To supplement this, future research could conduct user studies, either by employing casual readers or experienced Wikipedians, to qualitatively investigate the differences between protected and unprotected content.

Furthermore, our scope is limited to the English Wikipedia.
While protection levels, \rfpp\ archives, quality levels, and their predictors vary across language (see Section~\ref{sec:pp}), our experiments, code, and data can still provide a framework and serve as a baseline for the analyses of other Wikipedia versions.
As community dynamics and policies vary across languages (e.g., the Portuguese Wikipedia prohibits anonymous edits entirely), the effects of page protection on article quality could also differ~\cite{Johnson2022}.

\section{Conclusion}
Page protection is a core policy of Wikipedia, enforced by administrators to prevent content from harm by limiting contributions.
Although this mechanism protects the online encyclopedia from ``evil,'' such as vandalism, it can also prevent ``good'' by hindering article development.
In this work, we aim to assess how page protections on the English Wikipedia affect article quality, as measured by structural features of articles.
We characterize protections and user requests for protection for over a decade of data and provide, to the best of our knowledge, the first quantitative assessment of the effect of page protection on quality.
Using a quasi-experimental study, we find differential effects:
As high-quality articles benefit, low-quality ones may diminish in quality following protection.
Furthermore, effects vary across different topics.
These findings indicate that protecting high-quality articles may present a low risk, but protecting low-quality articles on certain topics requires caution.
Overall, our findings shed light on one of Wikipedia's most important content moderation tools and inform improvements to administrative processes on the Web.

\bibliographystyle{ACM-Reference-Format}
\bibliography{ref}
\newpage
\appendix
\section{Appendix: Data}
\label{app:data}

\xhdr{Missing data}
We retrieve \rfpp s starting from October 2012.
However, due to an error in the archiving on the English Wikipedia, October 2013 is missing from the archives and our dataset.

\begin{figure}[h]
	\includegraphics[width=\columnwidth,clip,trim={0 0 4cm 0},]{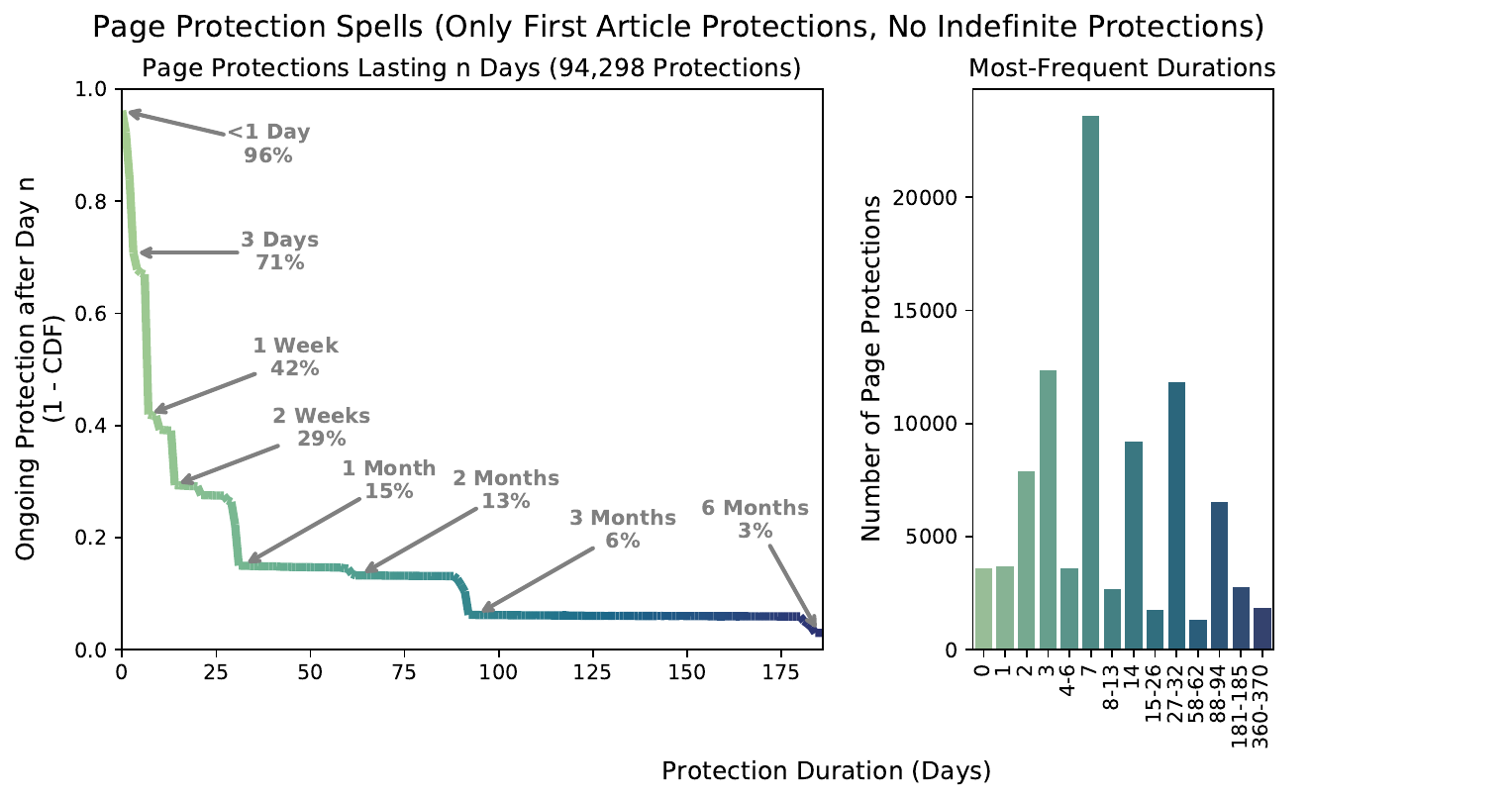}
	\caption{\textbf{Statistics of first article protections.}
		Duration distribution of the $\mathbf{94{,}298}$ first article protections does not differ considerably from our full dataset (Fig.~\ref{fig:pp_data_overview}). 
      Most protections are enforced for a week, but three-day, two-week, one-month, and three-month protections also occur regularly.
  } 
	\label{fig:pp_data_firstprotections}
\end{figure}

\xhdr{First protections}
We show statistics about first protections in Fig.~\ref{fig:pp_data_firstprotections} ($48\%$ of all protections).
Protection duration does not vary considerably between first and all article protections.

\begin{figure}[h]
	\includegraphics[width=\linewidth]{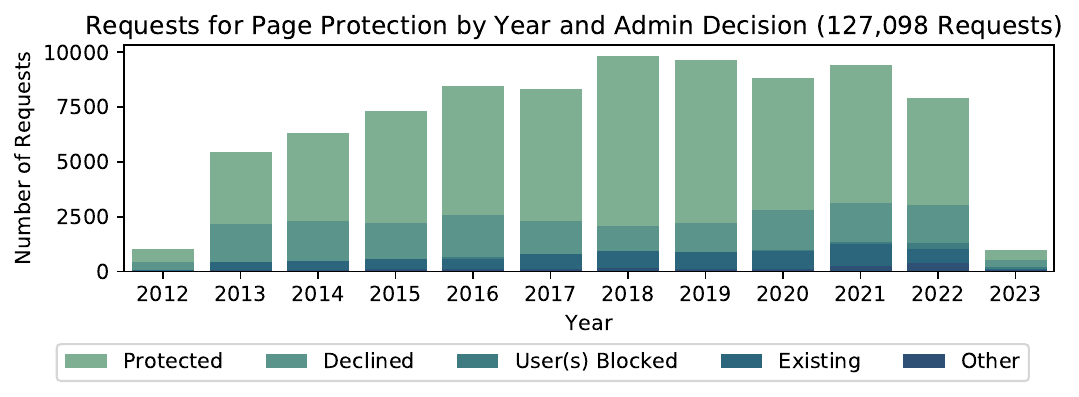}
	\caption{\textbf{Requests by year and decision.} 
		Admins enforce protection in response to most protection requests ($\mathbf{65.7\%}$), while $\mathbf{20.2\%}$ are declined, $\mathbf{6.4\%}$ lead to user interventions (e.g., blocking), $\mathbf{6.3\%}$ inform the user of existing protection, and $\mathbf{1.3\%}$ contain other responses (e.g., withdrawal of the \rfpp).} 
	\label{fig:pp_rfpp_overview}
\end{figure}

\xhdr{Requests over the years}
We visualize admin decisions to \rfpp s across years in Fig.~\ref{fig:pp_rfpp_overview}. 
Request numbers are relatively stable, with admin decisions in similar proportions over the years.

\begin{figure}[h]
    \centering
    \includegraphics[width=\columnwidth]{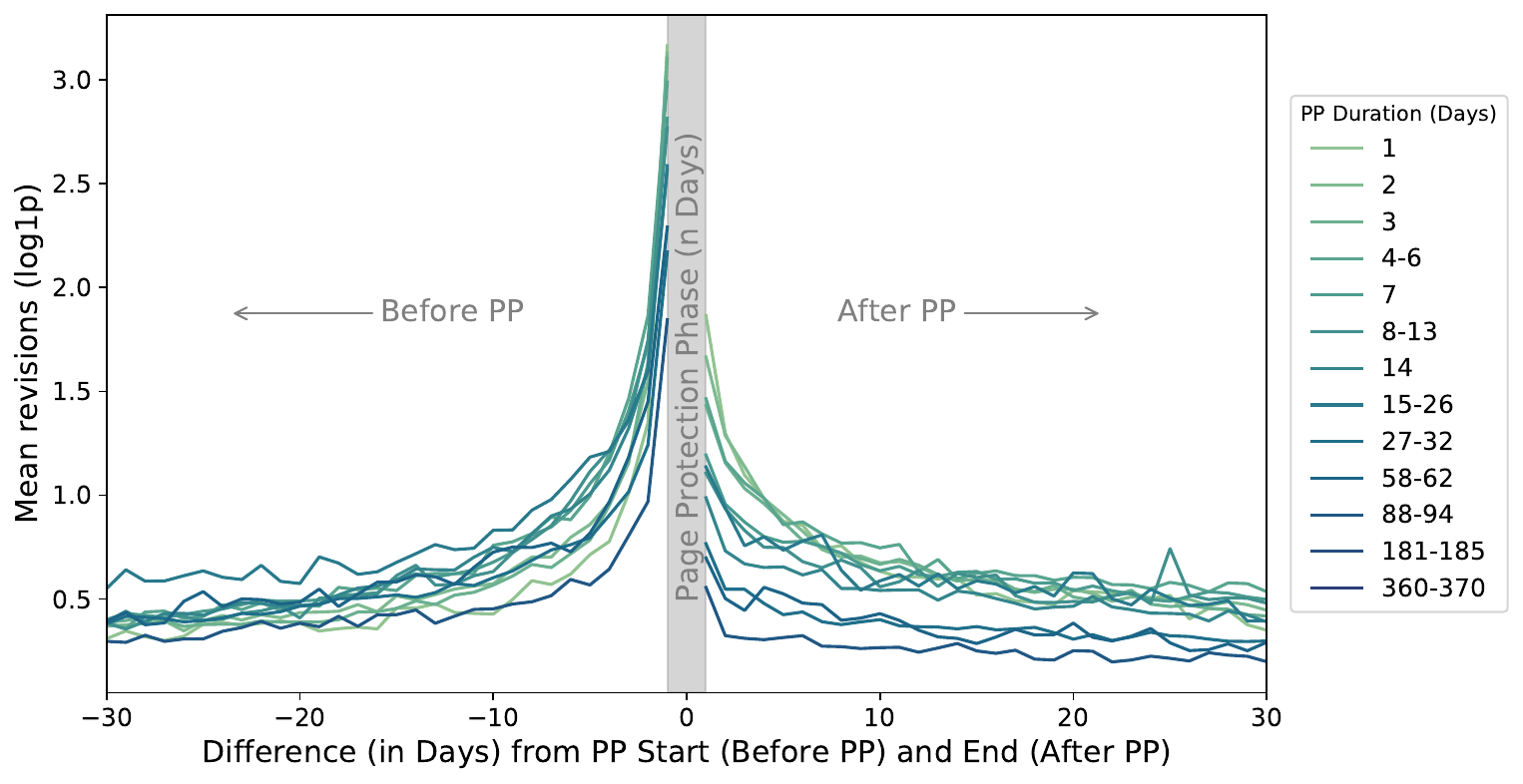}
    \caption{\textbf{Page protections by duration.} 
    We find that editing patterns suggest no substantial decreases after the end of protections, regardless of protection duration.}
    \label{fig:pp_durations_post}
\end{figure}

\xhdr{Editing Activity after Protection Ends}
We illustrate edits prior to and following protection in Fig.~\ref{fig:pp_durations_post} and observe that lifting page protection does not considerably increase edits.

\newpage
\section{Appendix: Methods}
\label{app:methods}
\xhdr{More on treatment and control groups}
We compare articles with an accepted protection request to those that were declined protection (Section~\ref{sec:experiment}).
We accept that the act of requesting article protection (even in the case of a declined request) can already be considered a treatment, as it possibly attracts the attention of admins and other experienced editors, who may perform minor janitorial actions on the page (e.g., selected reverts).
Note that we thus do not have a ``true control'' in our experiment, but instead use a ``quasi control'' in the form of declined \rfpp s.
The comparison of treatment of interest (i.e., article protection) with other ``baseline treatments'' (i.e., declined \rfpp s) is considered valid in designs where there is no true randomized control group~\cite{hernan2016using}.
As in our case, activity patterns on non-contentious articles may not be comparable to the increased attention that pages subject to a \rfpp\ receive and might introduce additional confounders.

\newpage
\section{Appendix: Results}
\label{app:results}

\begin{figure}[h]
	\centering
	\begin{subfigure}[h]{\columnwidth}
		\includegraphics[trim={0 0 0 .65cm},clip, width=\columnwidth]{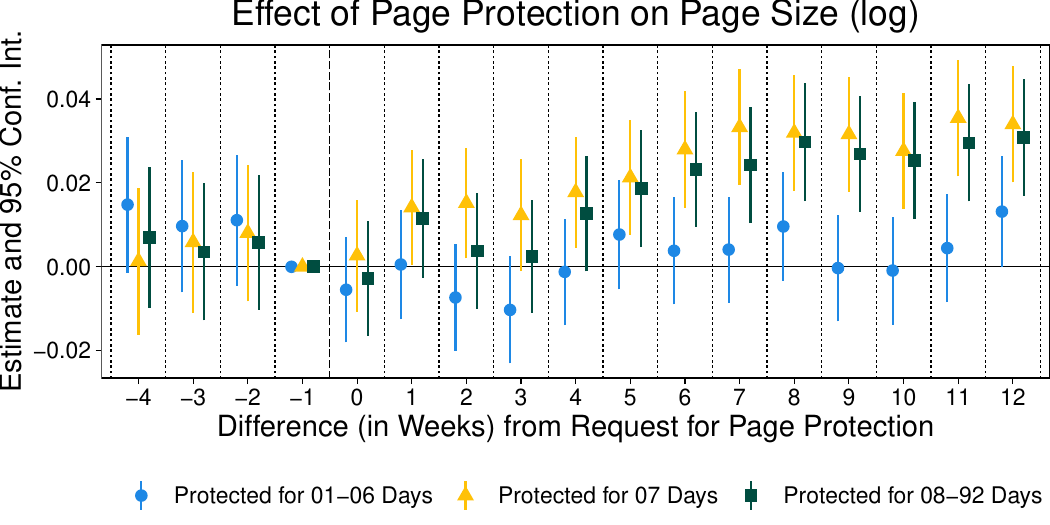}
		\subcaption{Effects for all Articles.}
			\label{fig:effect_size_all}
	\end{subfigure}
 
	\begin{subfigure}[h]{\columnwidth}
		\includegraphics[trim={0 0 0 .65cm},clip,width=\columnwidth]{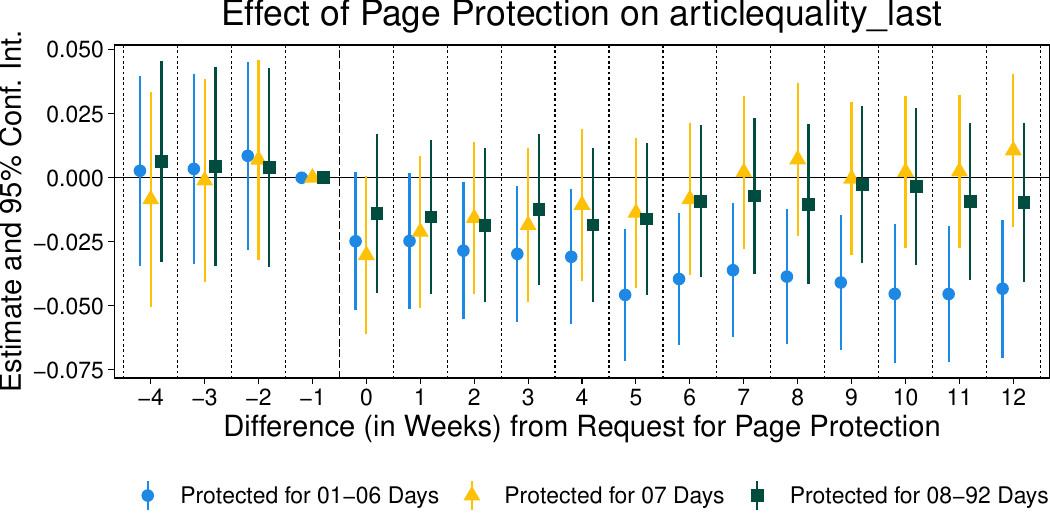}
		\subcaption{Effects for shorter articles.}
        \label{fig:effect_size_short}
	\end{subfigure}
 
	\begin{subfigure}[h]{\columnwidth}
		\includegraphics[trim={0 0 0 .65cm},clip,width=\columnwidth]{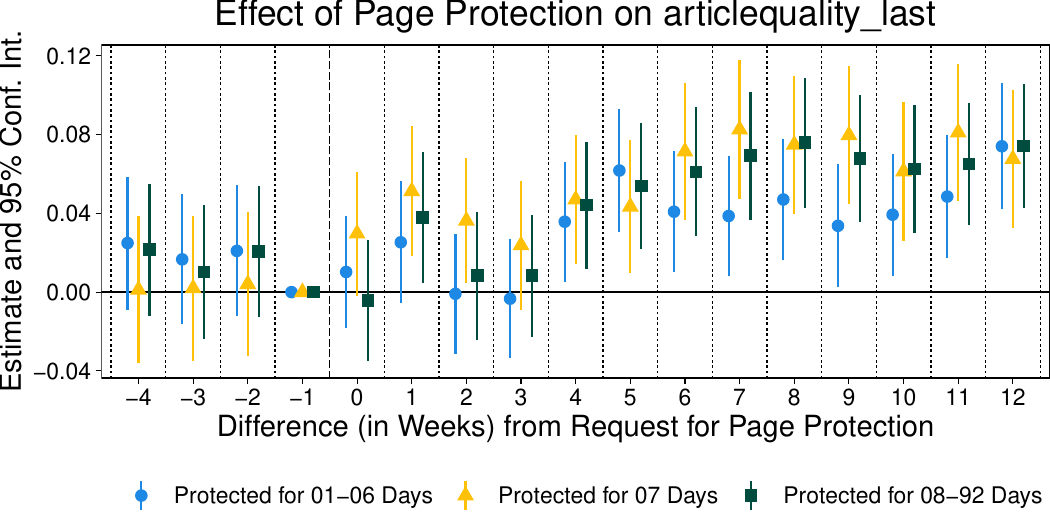}
 	\subcaption{Effects for longer articles.}
        \label{fig:effect_size_long}
	\end{subfigure}
	\caption{\textbf{Difference in article length after protection.}
		We fit dynamic difference-in-differences models estimating effect on article length (log scale, z-score standardized) on the matched dataset.
		Initial investigations on the matched dataset (a) indicate only minor differences in length between declined \rfpp\ and protected articles right after the request, but length rises to levels significantly higher than articles without protection.
		In (b) and (c), we fit the same model on small and large articles, and detect heterogeneous effects.
	}
	\label{fig:effect_size}
\end{figure}

\begin{figure}[h]
    \centering
    \includegraphics[width=\columnwidth]{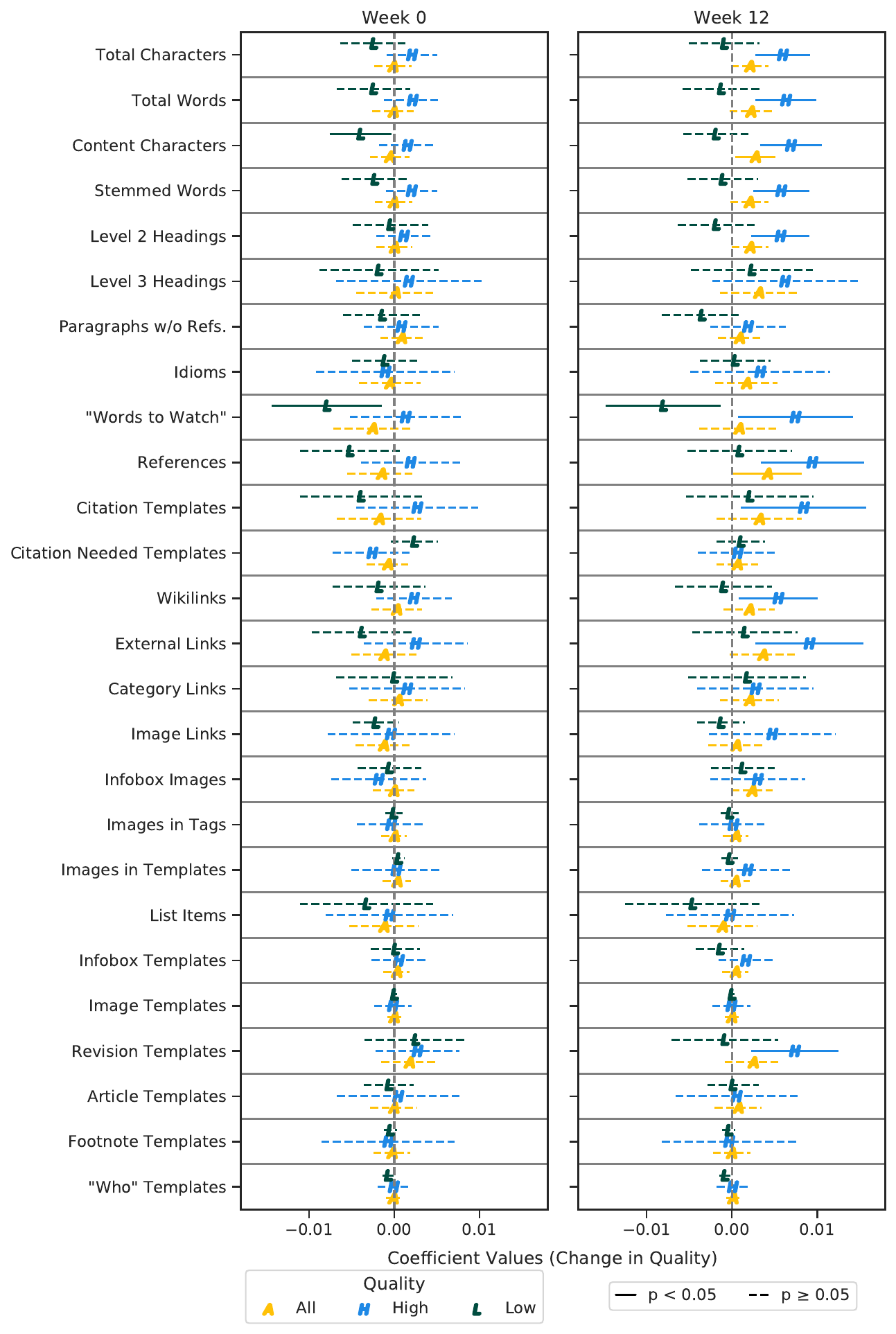}
    \caption{Coefficients for the constituents of article quality.
    Through binary difference\hyp{in\hyp{differences}} regression, we find that for our experiment, improvements (or decreases) in the ORES article quality metric are also driven by multiple other features besides article length.
    }
    \label{fig:aq_all_coefs}
\end{figure}

This appendix includes additional figures which were omitted from Section~\ref{sec:results} due to page limits.
First, we show all coefficients for the constituents of article quality in Figure~\ref{fig:aq_all_coefs}.
Next, we combine all relevant figures for our analysis of article length in Figure~\ref{fig:effect_size}.

\end{document}

%% file: dlabmacros.tex
\usepackage{hyphenat}
\usepackage{xspace}

\usepackage{amsthm}
\theoremstyle{plain}

\newcommand{\xhdr}[1]{\vspace{0.7mm}\noindent{{\bf #1.}}}

\newcommand{\textcite}[1]{\citeauthor{#1} \shortcite{#1}}

\newcommand{\hide}[1]{}

\makeatletter
\newcommand{\iffont}[2]{\ifthenelse{\equal{\f@family}{#1}}{#2}{}}
\makeatother

\iffont{ptm}{
  \usepackage{mathptmx}

  \DeclareSymbolFont{greek}{OML}{cmm}{m}{n}
  \DeclareMathSymbol{\alpha}{\mathalpha}{greek}{"0B}
  \DeclareMathSymbol{\beta}{\mathalpha}{greek}{"0C}
  \DeclareMathSymbol{\gamma}{\mathalpha}{greek}{"0D}
  \DeclareMathSymbol{\delta}{\mathalpha}{greek}{"0E}
  \DeclareMathSymbol{\epsilon}{\mathalpha}{greek}{"0F}
  \DeclareMathSymbol{\zeta}{\mathalpha}{greek}{"10}
  \DeclareMathSymbol{\eta}{\mathalpha}{greek}{"11}
  \DeclareMathSymbol{\theta}{\mathalpha}{greek}{"12}
  \DeclareMathSymbol{\iota}{\mathalpha}{greek}{"13}
  \DeclareMathSymbol{\kappa}{\mathalpha}{greek}{"14}
  \DeclareMathSymbol{\lambda}{\mathalpha}{greek}{"15}
  \DeclareMathSymbol{\mu}{\mathalpha}{greek}{"16}
  \DeclareMathSymbol{\nu}{\mathalpha}{greek}{"17}
  \DeclareMathSymbol{\xi}{\mathalpha}{greek}{"18}
  \DeclareMathSymbol{\pi}{\mathalpha}{greek}{"19}
  \DeclareMathSymbol{\rho}{\mathalpha}{greek}{"1A}
  \DeclareMathSymbol{\sigma}{\mathalpha}{greek}{"1B}
  \DeclareMathSymbol{\tau}{\mathalpha}{greek}{"1C}
  \DeclareMathSymbol{\upsilon}{\mathalpha}{greek}{"1D}
  \DeclareMathSymbol{\phi}{\mathalpha}{greek}{"1E}
  \DeclareMathSymbol{\chi}{\mathalpha}{greek}{"1F}
  \DeclareMathSymbol{\psi}{\mathalpha}{greek}{"20}
  \DeclareMathSymbol{\omega}{\mathalpha}{greek}{"21}
  \DeclareMathSymbol{\varepsilon}{\mathalpha}{greek}{"22}
  \DeclareMathSymbol{\vartheta}{\mathalpha}{greek}{"23}
  \DeclareMathSymbol{\varpi}{\mathalpha}{greek}{"24}
  \DeclareMathSymbol{\varrho}{\mathalpha}{greek}{"25}
  \DeclareMathSymbol{\varsigma}{\mathalpha}{greek}{"26}
  \DeclareMathSymbol{\varphi}{\mathalpha}{greek}{"27}
  \DeclareSymbolFont{otone}{OT1}{cmr}{m}{n}
  \DeclareMathSymbol{\Gamma}{\mathalpha}{otone}{0}
  \DeclareMathSymbol{\Delta}{\mathalpha}{otone}{1}
  \DeclareMathSymbol{\Theta}{\mathalpha}{otone}{2}
  \DeclareMathSymbol{\Lambda}{\mathalpha}{otone}{3}
  \DeclareMathSymbol{\Xi}{\mathalpha}{otone}{4}
  \DeclareMathSymbol{\Pi}{\mathalpha}{otone}{5}
  \DeclareMathSymbol{\Sigma}{\mathalpha}{otone}{6}
  \DeclareMathSymbol{\Upsilon}{\mathalpha}{otone}{7}
  \DeclareMathSymbol{\Phi}{\mathalpha}{otone}{8}
  \DeclareMathSymbol{\Psi}{\mathalpha}{otone}{9}
  \DeclareMathSymbol{\Omega}{\mathalpha}{otone}{10}
  \DeclareSymbolFont{syms}{OML}{cmm}{m}{it}
  \DeclareMathSymbol{\partial}{\mathord}{syms}{"40}
  \DeclareMathAlphabet{\mathbold}{OML}{cmm}{b}{it}
  \DeclareSymbolFont{largesymbols}{OMX}{cmex}{m}{n}

}